\documentclass[journal]{IEEEtran}

\usepackage{amsmath,amssymb}
\usepackage{graphicx}
\usepackage[hidelinks]{hyperref}
\usepackage{cite}

\begin{document}

\title{Turbulence Strength Estimation from Multiplexed Orbital Angular Momentum Beams Using Multimode Fibre Speckle}

\author{Warwick~G.~C.~Brown and Mitchell~A.~Cox,~\IEEEmembership{Senior Member,~IEEE}%
  \thanks{The authors are with the School of Electrical and Information Engineering, University of the Witwatersrand, Johannesburg 2050, South Africa. Corresponding author: M. A. Cox (e-mail: mitchell.cox@wits.ac.za).}}

\markboth{IEEE Photonics Technology Letters}%
{Brown and Cox: Turbulence Strength Estimation Using Multimode Fibre Speckle}

\maketitle

\begin{abstract}
The strength of atmospheric turbulence governs how much the modes of a free-space optical link mix together. It is usually measured with a separate instrument or with a Gaussian probe beam, neither of which uses the beams that a structured-light link already carries. Here we show that the turbulence strength, expressed as the Fried parameter, can be estimated directly from multiplexed orbital angular momentum (OAM) beams without knowing which modes are present. The beams pass through a passive multimode fibre that turns them into speckle on a camera, and a trained weighted sum of the camera pixels (known as an extreme learning machine), averaged over a few tens of frames, gives the Fried parameter. In the laboratory with emulated turbulence, superpositions of up to eleven OAM modes give an estimate as accurate as that from a Gaussian probe beam sent through the same fibre. In all but weak turbulence the estimate is also more accurate than an ideal Strehl meter or an ideal tip/tilt sensor, providing an efficient way to estimate turbulence strength in mode-division multiplexed systems.
\end{abstract}

\begin{IEEEkeywords}
Atmospheric turbulence, extreme learning machines, free-space optical communication, multimode fibre, orbital angular momentum.
\end{IEEEkeywords}

\IEEEpeerreviewmaketitle

\section{Introduction}
\IEEEPARstart{M}{ode} division multiplexing with orbital angular momentum (OAM) modes, or other forms of structured light, multiplies the capacity of a free-space optical link \cite{trichili2020roadmap,cox2021structured}. Atmospheric turbulence distorts these modes and causes crosstalk between them \cite{cox2019resilience,klug2021orbital}. How much the modes mix is governed by the strength of the turbulence, and a link is therefore characterised by measuring that strength.

The established ways of measuring it rely on something other than the beams that carry the data. Most transmit a Gaussian probe beam and record its scintillation index, its beam wander or angle of arrival, or the size of its seeing disc \cite{andrews2005laser,cox2021structured,tokovinin2002dimm}, and a link carrying OAM modes must send this probe in addition to its data. Others use a separate instrument, such as a scintillometer \cite{kleissl2008large}, a differential image motion monitor \cite{tokovinin2002dimm}, or a wavefront sensor \cite{cox2019wavefront}. Machine learning on camera images needs neither. Networks sort the turbulence into a few widely spaced levels \cite{yin2019experimental,jiang2025machine}, and one simulation study estimates the strength as a continuous value from a single frame \cite{sun2026single}. A single known OAM mode has also been used underwater, through the wander of its centroid \cite{omalley2026statistical} or a network that classifies the strength from its image \cite{omalley2026deep}. In each of these studies the beam sent is known in advance.

A link that carries data offers neither a probe nor a known beam. It sends superpositions of modes that keep changing, and the centroid of such a beam moves with the modes being sent as well as with the turbulence. In this letter we estimate the turbulence strength from simultaneously multiplexed OAM beams in the laboratory, a step left as future work in \cite{sun2026single}. The method is an optical extreme learning machine \cite{rawlings2026exploiting,huang2006extreme,pierangeli2021photonic}, which is a network whose single hidden layer is random and fixed, so that only its output weights are trained. Here a 3~m OM1 multimode fibre acts as the hidden layer by scrambling the beam into speckle, the square-law detection of a camera supplies the nonlinearity, and the trained output, which we call the readout, is a weighted sum of the camera pixels. Superpositions of up to eleven OAM modes are sent through emulated turbulence, and the readout, blind to the modes present, gives the Fried parameter once averaged over a few tens of frames. From 25 frames the median error is 5.8\%, against 7.0\% for an ideal Strehl meter and 8.0\% for an ideal tip/tilt sensor.

\section{Principle}

Turbulence strength is typically described by the Fried parameter $r_0$, which is the distance across which the phase of the distorted wavefront stays correlated, and by $D/r_0$, which compares the size of the beam $D$ with $r_0$ \cite{andrews2005laser,klug2021orbital}. A smaller $r_0$ means stronger turbulence. We apply the turbulence as a phase screen $\psi$ that multiplies the beam by $\exp(i\psi)$. Each frame is one camera image, for which the transmitter sends either a Gaussian beam $E_0$ or a superposition of 1 to 11 Laguerre--Gaussian modes, $E_\mathrm{data} = \sum_{\ell=-5}^{5}\sqrt{p_\ell}\,E_\ell$, where the azimuthal index $\ell$ sets the OAM of each mode and $p_\ell$ is the mode's share of the power \cite{cox2021structured}. For the Gaussian beam, the Strehl ratio of a single screen is the on-axis intensity of the focused beam divided by that without turbulence \cite{cox2019resilience},
\begin{equation}
S = \frac{\left|\int E_0\, \exp(i\psi)\,\mathrm{d}A\right|^2}{\left|\int E_0\,\mathrm{d}A\right|^2}.
\label{eq:strehl}
\end{equation}
Its average falls as $D/r_0$ rises, but screens of the same turbulence strength differ from one another by chance. For example, at $D/r_0 = 0.63$ the middle 80\% of screens have a Strehl ratio anywhere between 0.24 and 0.83. The strength cannot therefore be read from the Strehl ratio of one screen, and the frames of different screens must be averaged.

The multimode fibre guides about 2600 modes, which interfere at its output to form a speckle pattern. The fibre thus scrambles the field through a transmission matrix $\mathbf{T}$ that is fixed and never trained. If $\mathbf{e}$ is the field leaving the screen and $\mathbf{t}_k$ is the row of $\mathbf{T}$ belonging to pixel $k$, with the optics before the fibre included in $\mathbf{T}$, then each pixel measures
\begin{equation}
h_k = \left|\mathbf{t}_k^{\mathsf T}\mathbf{e}\right|^2 = \mathbf{e}^\dagger\mathbf{Q}_k\mathbf{e},\qquad \mathbf{Q}_k = \mathbf{t}_k^{*}\mathbf{t}_k^{\mathsf T},
\label{eq:proj}
\end{equation}
which is quadratic in the field. The Strehl ratio is quadratic in that same field, so it can be written as a weighted sum of the pixels,
\begin{equation}
S = \frac{\mathbf{e}^\dagger\left(\mathbf{u}\mathbf{u}^\dagger\right)\mathbf{e}}{\left|\mathbf{u}^\dagger\mathbf{e}_0\right|^2} = \sum_k w_k h_k ,
\label{eq:linear}
\end{equation}
where $\mathbf{u}$ is a vector of ones that sums the field and $\mathbf{e}_0$ is the field without turbulence. The second equality requires the pixels to supply enough independent quadratic forms. A quadratic form of a field with $n$ dimensions has $n^2$ real degrees of freedom, so the $10^4$ pixel values used here can supply an arbitrary one only if the distorted field occupies at most approximately 100 dimensions. We expect this of the turbulence strengths considered, and test it in Section~\ref{sec:results}. The weights, together with a constant offset, are found by ridge regression on the standardised pixel values of frames where the answer is known. The only nonlinearity needed is therefore the square-law detection of the camera. Note that the phase itself is not quadratic in the field and cannot be read in this way.

The Fried parameter, however, is not a property of a single frame. When \eqref{eq:proj} is averaged over screens, the products of the field at two points a distance $\Delta$ apart are reduced by the coherence of the turbulence, $\exp[-3.44(\Delta/r_0)^{5/3}]$ \cite{tokovinin2002dimm}. The mean of every pixel, and of any weighted sum of pixels, is thus a smooth function of $r_0$ for any beam sent. A Strehl meter corresponds to one fixed choice of weights, namely those of \eqref{eq:linear}, so averaging Strehl ratios could at best match it. We therefore train a second readout of the same form on $D/r_0$ directly, which leaves the regression free to choose weights that respond more strongly to the turbulence. Its outputs $\hat y_i$ are averaged over $N$ frames of different screens,
\begin{equation}
\widehat{D/r_0} = g^{-1}\!\left(\frac{1}{N}\sum_{i=1}^{N}\hat y_i\right), \qquad \hat y_i = \sum_k w_k h_{k,i},
\label{eq:average}
\end{equation}
where the calibration curve $g$ is the average output of the readout at each known $D/r_0$. Dividing $D$ by the estimate gives $r_0$. The error of the estimate is approximately $\sigma/(|g'|\sqrt{N})$, where $\sigma$ is the spread of $\hat y_i$ between screens of one strength and $g'$ is the slope of the calibration curve.

\section{Experimental Setup}

\begin{figure}[t]
    \centering
    \includegraphics[width=0.9\linewidth]{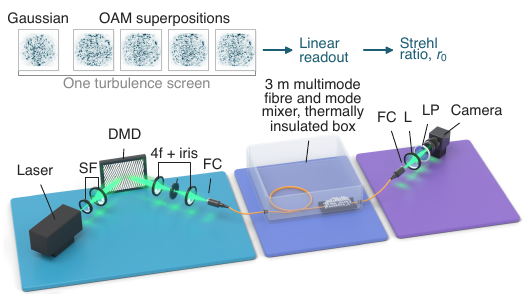}
    \caption{The experimental setup. 
    The laser is spatially filtered (SF) and shaped by a digital micromirror device (DMD).
    A 4f system with an iris selects the first diffraction order, and fibre couplers (FC) launch it into and collect it from the multimode fibre. The output passes through a lens (L) and a linear polariser (LP) to a camera.}
    \label{fig:setup}
\end{figure}

The setup is shown in Fig.~\ref{fig:setup}. A 532~nm laser is shaped by a digital micromirror device, which writes the mode and the phase screen onto the beam in a single hologram \cite{cox2021dmd}. The screen is thus applied in the near field, and the beam is then focused onto the multimode fibre, so the fibre sits in the far field of the screen. It passes through a mode mixer (the fibre is weighed down with many steel balls), which couples light between its modes, and it is kept in a thermally insulated box, which slows the drift of the speckle. A camera records the speckle through a linear polariser with an exposure of 2~ms. This exposure keeps the camera within its linear range, as \eqref{eq:linear} requires, whereas in \cite{pierangeli2021photonic,rawlings2026exploiting} the camera is saturated on purpose to add nonlinearity.
The turbulence is emulated with Kolmogorov phase screens generated by the Fourier transform method with added subharmonics \cite{lane1992simulation}. The screens are thin, in that they change only the phase of the beam, so there is no scintillation and scintillation methods are not compared. The tilt of each screen is retained, which the tip/tilt comparison below relies on. Because we generate every screen, its Strehl ratio from \eqref{eq:strehl} and its $r_0$ are known exactly, which is not possible outdoors. Three kinds of screen are shown in random order: no screen at all, strengths sampled continuously from $D/r_0 = 0.05$ to 3, and ten fixed strengths from $D/r_0 = 0.061$ to 2.35. The beam size is $D = 2w_0 = 1$~mm, where $w_0$ is the beam waist radius, so over the ten fixed strengths $r_0$ moves from 16 to 0.43~mm and the average Strehl ratio falls from 0.99 to 0.10. Frames are captured in bursts of one Gaussian frame followed by four OAM frames, with a new screen for every burst. In an OAM frame the number of modes, the modes themselves and their power shares (at least 5\% each) are random. The modes are unmodulated and added in phase. About 2.3 million frames are used, captured at approximately 16.5 frames per second.

Each 600 by 600 pixel image is cropped to 450 by 450 pixels and block averaged to 100 by 100. To counter the drift of the speckle, a block of 38 fixed reference patterns is displayed every 500 frames, and each image is divided pixel by pixel by the mean of the most recent block. These operations are linear in the pixel values, which keeps \eqref{eq:linear} valid. The speckle still decorrelates over a few hours, so the readouts are trained separately for every 2-hour window. Each window is split into four blocks of time, and every frame is estimated by readouts trained on the other three blocks, which hold 11\,000 Gaussian or 41\,000 OAM frames. No screen is therefore shared between training and testing. Three readouts are trained in this way. Two use the Gaussian frames, one for the Strehl ratio and one for $D/r_0$, and the third estimates $D/r_0$ from the OAM frames of all superpositions together. The calibration curve $g$ of each $D/r_0$ readout is built from its outputs on the continuous-strength frames, averaged in bins of $D/r_0$, and all errors in $r_0$ are reported on the fixed-strength frames.

We compare against three sensors that would each use a separate Gaussian beam. A Strehl meter places a detector at the focus of the beam to record how far the on-axis intensity falls \cite{cox2019resilience}, a tip/tilt sensor follows the beam wander or angle of arrival \cite{cox2021structured}, and a wavefront sensor measures the phase itself \cite{cox2019wavefront}. All three are ideal, in that they are computed from the known screens and so have no measurement noise. The ideal Strehl meter uses the Strehl ratio $S$ of each screen, the ideal tip/tilt sensor uses the squared tilt of the plane that best fits each screen, and the ideal wavefront sensor uses the phase variance that remains once that plane is removed. Every method is evaluated on the same screens and converted to $r_0$ using \eqref{eq:average} with its own calibration curve. The error of a set of $N$ frames, each from a different screen of one strength, is $|\hat r_0/r_0 - 1|$, and we report its median over the sets.

\section{Results and Discussion} \label{sec:results}

\begin{figure}[t]
    \centering
    \includegraphics[width=0.9\linewidth]{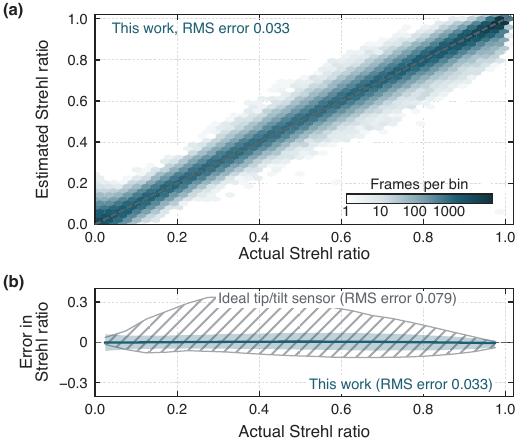}
    \caption{(a) Estimated against actual Strehl ratio for 372\,822 single Gaussian frames. The dashed line is perfect agreement.
    (b) Error against actual Strehl ratio for this work (median line with its 5--95\% band) and for an ideal tip/tilt sensor on the same screens (hatched, 5--95\% band).}
    \label{fig:strehl}
\end{figure}

The principle is first checked on single Gaussian frames in Fig.~\ref{fig:strehl}, where \eqref{eq:linear} predicts that the Strehl ratio can be read directly. The estimate follows the actual value across the whole range of turbulence, with a root-mean-square error of 0.033 and a correlation coefficient of 0.996, and 90\% of frames fall within $\pm 0.05$. This is consistent with the pixels supplying the quadratic form that \eqref{eq:linear} requires. An ideal tip/tilt sensor gives an error of 0.079 when the tilt of each screen is mapped to the Strehl ratio by a fitted curve, meaning that the readout captures more of the distortion than the tilt alone. The Strehl ratio, however, is defined for the Gaussian beam. The Fried parameter does not depend on the beam, which leads us to the main result.

\begin{figure*}[t]
    \centering
    \includegraphics[width=0.88\linewidth]{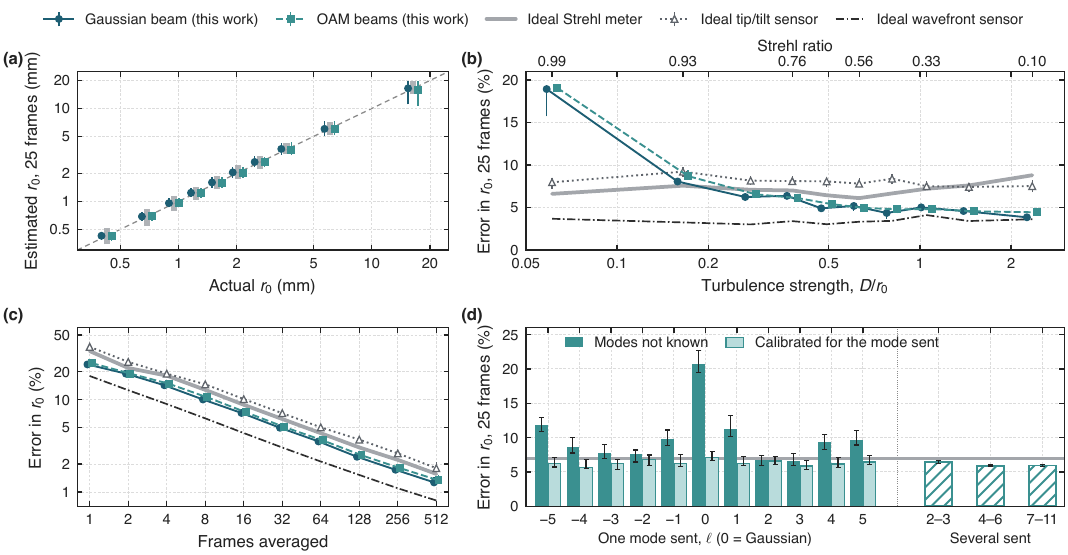}
    \caption{Fried parameter $r_0$ from averaged Gaussian frames and from OAM superposition frames whose modes are not known to the readout, compared with ideal sensors on the same screens.
    (a) Estimated against actual $r_0$ from 25 frames (median and 5--95\% range of the sets, with the ideal Strehl meter in grey).
    (b) Error from 25 frames at each strength.
    (c) Error against the number of frames averaged, over all ten strengths.
    (d) Error from 25 frames by what was sent, either one mode alone ($\ell = 0$ is the Gaussian) or several together (hatched). Dark bars are for the OAM readout and light bars for the same readout with its calibration curve rebuilt for the mode sent. The grey line is the ideal Strehl meter. Error bars in (b)--(d) show 95\% bootstrap confidence intervals.}
    \label{fig:r0}
\end{figure*}

The OAM superpositions estimate $r_0$ as accurately as a Gaussian beam does, and more accurately than an ideal Strehl meter or an ideal tip/tilt sensor. Fig.~\ref{fig:r0}(a) shows that the estimate from 25 frames follows the actual $r_0$ over the full range for both kinds of frame. Across all ten strengths the median error is 5.8\% for the OAM superpositions and 5.7\% for the Gaussian beam, whereas an ideal Strehl meter achieves 7.0\% and an ideal tip/tilt sensor 8.0\% on the same screens.

This advantage depends on the strength of the turbulence, as shown in Fig.~\ref{fig:r0}(b). From $D/r_0 = 0.28$ upwards, where the Strehl ratio is 0.84 or lower, the error of the readout lies between 4 and 7\%, while that of the two ideal sensors lies between 6 and 9\%. The gap to the Strehl meter widens in stronger turbulence, which we attribute to the Strehl ratio flattening towards zero, so that its calibration curve loses its slope. At $D/r_0 = 0.17$ the readout and the two sensors are comparable. At the weakest strength of $D/r_0 = 0.06$ the readout is worse, at 19\% against 7 to 8\%. Here the beam experiences almost no turbulence, so the speckle barely changes and the noise of the measurement dominates, whereas the ideal sensors have no noise. An ideal wavefront sensor achieves 3 to 4\% at every strength. The readout therefore sits between these two sensors and a wavefront sensor.

Fewer frames are also required, as seen in Fig.~\ref{fig:r0}(c). An error of 10\% is reached with 8 to 9 frames, compared with about 13 for the ideal Strehl meter and 16 for the ideal tip/tilt sensor, and an error of 5\% with 32 frames compared with 48 and 64. The error continues to fall as approximately $1/\sqrt{N}$, reaching 1.8\% at 256 frames.

The readout does not need to know which modes are present, provided that it is calibrated on a similar mix. Fig.~\ref{fig:r0}(d) splits the error by what was sent, with each set now holding frames of one kind, so an offset particular to that kind no longer averages away. Superpositions of 2 to 3, 4 to 6 and 7 to 11 modes achieve 6.5, 5.9 and 5.9\% respectively. Single modes sent alone range from 7\% for $|\ell| = 2$ and 3 up to 11.8\% for the other OAM modes, and a Gaussian frame sent through the OAM readout reaches 20.7\%. This is expected, because the calibration curve $g$ depends on the beam and is built from what the readout sees most, namely superpositions. We attribute the spread to superpositions behaving on average like a mode in the middle of the set, whereas higher $|\ell|$ reacts more strongly to turbulence and lower $|\ell|$ less so \cite{klug2021orbital}. If one keeps the weights and rebuilds $g$ from frames of the single mode that is sent, every mode comes down to between 5.7 and 7.1\%, level with or below the ideal Strehl meter.

Some practical points must be emphasised. Each frame averaged here comes from a different screen, whereas on a real link the frames must be spaced far enough apart for the air to change. The readout must be trained against an independent measurement of the turbulence strength, and here it was refitted every 2 hours to follow the drift of the fibre. The modes were also unmodulated and added in phase. Finally, we note that the same arrangement of multimode fibre, camera and linear readout has been shown to classify OAM modes through turbulence \cite{rawlings2026exploiting}, so the method could one day share a receiver with a demultiplexer.

\section{Conclusion}
We have shown in the laboratory that the turbulence strength of a free-space optical link can be estimated from the multiplexed OAM beams that the link already carries, without a Gaussian probe and with the readout blind to the modes present. A multimode fibre and a camera, read by a trained weighted sum of the pixels, give the Fried parameter as accurately as a Gaussian probe sent through the same fibre, and in all but weak turbulence more accurately than an ideal Strehl meter or an ideal tip/tilt sensor. We attribute this advantage to the speckle offering the readout many quadratic measurements of the field, where a Strehl meter has only one. The method must still be tested on a real path, with scintillation and with modulated data. Should it hold there, a structured-light link could monitor its own turbulence using only the beams that it sends.

\ifCLASSOPTIONcaptionsoff
  \newpage
\fi

\bibliographystyle{IEEEtran}
\bibliography{references}

\end{document}